\documentclass[twocolumn,prl,showpacs,preprintnumbers,amsmath,amssymb,superscriptaddress]{revtex4}
\usepackage{graphicx,color}
\usepackage{amssymb}
\usepackage{amsmath}
\usepackage{latexsym}
\usepackage{amsxtra}
\usepackage{amscd}
\usepackage{dcolumn}
\usepackage{bm}

\usepackage{hyperref}

\begin{document}

\title{Probing elastic and inelastic breakup contributions to intermediate-energy two-proton removal reactions}

\author{K.~Wimmer}

\author{D.~Bazin}
\affiliation{National Superconducting Cyclotron Laboratory, Michigan State University, East Lansing, Michigan 48824, USA}

\author{A.~Gade}
\affiliation{National Superconducting Cyclotron Laboratory, Michigan State University, East Lansing, Michigan 48824, USA}
\affiliation{Department of Physics and Astronomy, Michigan State University, East Lansing, Michigan 48824, USA}

\author{J.~A.~Tostevin}
\affiliation{National Superconducting Cyclotron Laboratory, Michigan State University, East Lansing, Michigan 48824, USA}
\affiliation{Department of Physics, University of Surrey, Guildford, Surrey GU2 7XH, United Kingdom}

\author{T.~Baugher}
\affiliation{National Superconducting Cyclotron Laboratory, Michigan State University, East Lansing, Michigan 48824, USA}
\affiliation{Department of Physics and Astronomy, Michigan State University, East Lansing, Michigan 48824, USA}

\author{Z.~Chajecki}
\affiliation{National Superconducting Cyclotron Laboratory, Michigan State University, East Lansing, Michigan 48824, USA}

\author{D.~Coupland}
\affiliation{National Superconducting Cyclotron Laboratory, Michigan State University, East Lansing, Michigan 48824, USA}
\affiliation{Department of Physics and Astronomy, Michigan State University, East Lansing, Michigan 48824, USA}

\author{M.~A.~Famiano}
\affiliation{Department of Physics, Western Michigan University, Kalamazoo, Michigan 49008, USA}

\author{T.~K.~Ghosh}
\affiliation{Variable Energy Cyclotron Centre, 1/AF Bidhannagar, Kolkata 700064, India}

\author{G.~F.~Grinyer}\altaffiliation[Present address ]{GANIL, CEA/DSM-CNRS/IN2P3, Bvd Henri Becquerel, 14076 Caen, France}
\affiliation{National Superconducting Cyclotron Laboratory, Michigan State University, East Lansing, Michigan 48824, USA}

\author{R.~Hodges}
\affiliation{National Superconducting Cyclotron Laboratory, Michigan State University, East Lansing, Michigan 48824, USA}
\affiliation{Department of Physics and Astronomy, Michigan State University, East Lansing, Michigan 48824, USA}

\author{M.~E.~Howard}
\affiliation{Department of Physics and Astronomy, Rutgers University, New Brunswick, New Jersey 08903, USA}

\author{M.~Kilburn}
\author{W.~G.~Lynch}
\affiliation{National Superconducting Cyclotron Laboratory, Michigan State University, East Lansing, Michigan 48824, USA}
\affiliation{Department of Physics and Astronomy, Michigan State University, East Lansing, Michigan 48824, USA}

\author{B.~Manning}
\affiliation{Department of Physics and Astronomy, Rutgers University, New Brunswick, New Jersey 08903, USA}

\author{K.~Meierbachtol}
\affiliation{National Superconducting Cyclotron Laboratory, Michigan State University, East Lansing, Michigan 48824, USA}
\affiliation{Department of Chemistry, Michigan State University, East Lansing, Michigan 48824, USA}

\author{P.~Quarterman}
\author{A.~Ratkiewicz}
\author{A.~Sanetullaev}
\author{S.~R.~Stroberg}
\affiliation{National Superconducting Cyclotron Laboratory, Michigan State University, East Lansing, Michigan 48824, USA}
\affiliation{Department of Physics and Astronomy, Michigan State University, East Lansing, Michigan 48824, USA}

\author{M.~B.~Tsang}
\affiliation{National Superconducting Cyclotron Laboratory, Michigan State University, East Lansing, Michigan 48824, USA}

\author{D.~Weisshaar}
\affiliation{National Superconducting Cyclotron Laboratory, Michigan State University, East Lansing, Michigan 48824, USA}

\author{J.~Winkelbauer}
\affiliation{National Superconducting Cyclotron Laboratory, Michigan State University, East Lansing, Michigan 48824, USA}
\affiliation{Department of Physics and Astronomy, Michigan State University, East Lansing, Michigan 48824, USA}

\author{R.~Winkler}
\affiliation{National Superconducting Cyclotron Laboratory, Michigan State University, East Lansing, Michigan 48824, USA}

\author{M.~Youngs}
\affiliation{National Superconducting Cyclotron Laboratory, Michigan State University, East Lansing, Michigan 48824, USA}
\affiliation{Department of Physics and Astronomy, Michigan State University, East Lansing, Michigan 48824, USA}

\begin{abstract}
The two-proton removal reaction from $^{28}$Mg projectiles has been 
studied at 93~MeV/u at the NSCL. First coincidence measurements of the
heavy $^{26}$Ne projectile residues, the removed protons and other
light charged particles enabled the relative cross sections from each
of the three possible elastic and inelastic proton removal mechanisms
to be determined. These more final-state-exclusive measurements are
key for further interrogation of these reaction mechanisms and use
of the reaction channel for quantitative spectroscopy of very neutron-rich 
nuclei. The relative and absolute yields of the three contributing 
mechanisms are compared to reaction model expectations - based on the 
use of eikonal dynamics and $sd$-shell-model structure amplitudes.
\end{abstract}

\maketitle
\section{Introduction}
In fast surface-grazing collisions with a light target nucleus the removal 
of two protons (neutrons) from an intermediate-energy neutron-rich (deficient) 
projectile beam has been shown to proceed as a {\it direct} reaction 
\cite{bazin03,tostevin04,yoneda06}. Both the measured distributions of 
the reaction cross section among the available bound final states of the 
projectile-like residues \cite{tostevin06,yoneda06} and the associated 
residue momentum distributions \cite{simpson09,simpson09b} are correctly 
predicted using an eikonal-model description of the reaction dynamics, 
that assumes sudden, single-step two-nucleon removal, plus shell-model 
two-nucleon structure information. The reaction mechanism, that in general 
populates several final states, thus gives access to spectroscopic 
information on states of some of the most neutron-rich (deficient) exotic 
species, e.g. \cite{gade07a,bastin07,gade07b}. The reaction also promises 
a rather unique and practical tool to probe the spatial correlations of 
the removed nucleons. It is these correlations that drive the sensitivity 
of the momentum distributions to the spins, $J$, of the final-states, 
allowing $J$-assignments of states of the projectile-like residues 
\cite{simpson09,simpson09b,wiedenho11}. Two-nucleon removal reactions  
thus offer a complementary method to study the structure of exotic nuclei 
at fragmentation facilities. Compared to Coulomb excitation, they populate 
a more diverse set of final-state spins $J^\pi$~\cite{michimasa06}.

In a one-nucleon knockout reaction the nucleon is removed in a peripheral 
collision of the projectile with the target nucleus. Events can result from
elastic (diffractive dissociation) or inelastic (stripping) collisions
between the nucleon and the target, which must be summed. The first precise 
measurements to quantify the relative importance of these two reaction 
mechanisms used the one-proton removal reaction from the weakly-bound 
proton-rich nuclei $^{9}$C and $^{8}$B \cite{bazin09}. Both the measured 
cross sections and the relative contributions from the stripping and 
diffraction events were in good agreement with the reaction calculations 
with their sudden, eikonal dynamics description.

It follows, in one-step two-nucleon removal, that three reaction mechanisms
may contribute to the cross section: (i) the inelastic removal of both
nucleons (stripping), (ii) the elastic removal of one nucleon and the 
inelastic removal of the second (diffraction-stripping), and (iii)
the elastic dissociation of both nucleons (diffraction). Both (i) and
(ii) involve energy transfer to and excitation of the target nucleus. 
Calculations based on eikonal reaction dynamics and $sd$-shell-model 
structure input \cite{tostevin04,tostevin06} can provide quantitative 
estimates of the yields due to these different mechanisms. In particular, 
the diffraction mechanisms (ii) and (iii), where at least one of the 
nucleons is removed by elastic interactions with the target, are 
predicted to contribute about 40~\% of the two-proton removal cross 
section, even for the case considered here which has well-bound protons 
with two-proton separation energy $S_\text{2p}$= 30.0 MeV~\cite{audi03}. 
The reaction model, at present, makes predictions for the cross sections 
from these three mechanisms that (a) are exclusive with respect to the 
final states of the heavy projectile-like residues, as is essential for 
spectroscopy of these products, but (b) are inclusive with respect to
the (complete) set of final states of the target and of the removed 
protons.

A more quantitative confrontation of these reaction model predictions, 
through measurements of these three contributing reaction mechanisms, is
of importance for the development of and confidence in the exploitation 
of these techniques. In this article we present the first such measurements 
to determine the relative importance of the three two-nucleon removal 
mechanisms defined above, now denoted str, diff-str and diff, via the
measurement also of light, charged reaction fragments. The new measurements 
are also able to characterize aspects of the correlations of the detected 
protons in the final state, which will be discussed elsewhere~\cite{2pcorrel}.

The inclusive $^{9}$Be($^{28}$Mg,$^{26}$Ne)X reaction is used at an 
intermediate energy of 93 MeV/u. The reaction calculations for this system, 
using the methods and inputs detailed in Ref.~\cite{tostevin06}, predict 
that the two-proton removal cross sections due to mechanisms (i)
and (ii) are $\sigma_{\text{str}}=1.70$~mb and $\sigma_{\text{diff-str}}
=1.13$~mb, respectively. The cross section for process (iii), the
elastic removal of both protons is small and was only estimated,
based on the calculated probability for one of the
nucleons to be diffracted, i.e. $\sigma_{\text{diff}} = \left[
\sigma_\text{diff-str} / (2\cdot\sigma_{\text{str}}) \right]^2\times
\sigma_{\text{str}}$, giving 0.19~mb for the present case. Thus the
measurement of this diffraction component (predicted to be only 
6.3~\% of the total yield) is a significant experimental challenge. 
The chosen reaction was studied previously in a $^{26}$Ne--$\gamma$ 
coincidence measurement~\cite{bazin03}, and was used to study the 
relative populations of the four observed $^{26}$Ne bound final states. 
These relative populations were in excellent agreement with the direct
two-proton removal mechanism predictions and the $sd$-shell-model 
spectroscopy, that is also used here \cite{tostevin06}. Furthermore, 
these calculations predict that the fractional contributions of the 
three removal mechanisms to each of the $^{26}$Ne final states are 
essentially the same. Thus, since this previous experiment was not 
designed to detect light charged particles in the final-state, it 
could not differentiate the contributions from the individual
reaction mechanisms, the subject of interest here.

\section{Experiment}
\begin{figure}[h]
\centering
\includegraphics[width=\columnwidth]{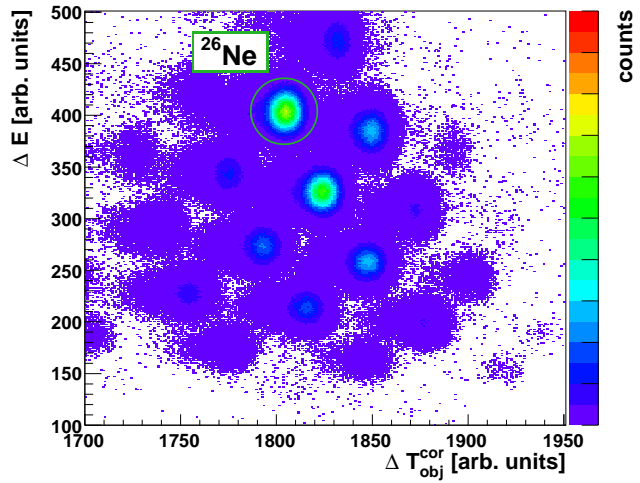}
\caption{(color online) Event-by-event particle identification spectrum of the reaction residues from the $^{28}$Mg secondary beam. Plotted is the energy loss versus the corrected time of flight measured between scintillators before and after the target. The circle indicates the $^{26}$Ne two-proton removal reaction residues of interest.}
\label{fig:pid}
\end{figure}

The $^{9}$Be($^{28}$Mg,$^{26}$Ne) reaction measurement was performed at the Coupled
Cyclotron Facility at NSCL. The $^{28}$Mg secondary beam with an
energy of 93~MeV/u was produced by projectile fragmentation of a 140
MeV/u $^{40}$Ar primary beam and selected using the A1900 fragment
separator~\cite{morrissey03}. The $^{9}$Be reaction target with a
thickness of 100~mg/cm$^2$ was placed at the target position of the
high-resolution S800 magnetic spectrograph~\cite{bazins80003}. The
incoming beam was identified event-by-event from the measurement of
the time-of-flight difference between two plastic timing scintillators
positioned before the target. The purity of $^{28}$Mg in the beam was 98.5~\% with
a rate on target of typically $5\times10^5$ particles/s. Two position
sensitive PPACs allowed for the correction of the momentum dispersion
in the incoming beam. The $^{26}$Ne reaction residues were identified
event-by-event by measuring the energy loss in the ionization chamber
in the S800 focal plane and the time of flight between scintillators
before and after the target, corrected for the trajectory of the ion, as shown in Fig.~\ref{fig:pid}.
The $^{26}$Ne energies and momenta were
reconstructed from the measured positions and angles in the S800 focal
plane using the position-sensitive CRDCs and ray-tracing. Protons
and other light charged particles were detected and identified in the
High Resolution Array (HiRA)~\cite{wallace07}. In the configuration used in the present study, this array consists of 17 $\Delta E - E$ telescopes made of 1.5~mm thick double-sided silicon strip detectors backed by 4~cm deep CsI crystals.

This allowed for the identification of protons with
kinetic energies larger than 15~MeV and other light charged particles such as deuterons and tritons. The polar angle coverage was from 9$^\circ$
to 56$^\circ$ in the laboratory system. The solid angle covered by the HiRA array and the geometric efficiency for the configuration used in the present study are shown in Fig.~\ref{fig:extr} (a) and (b), respectively.

\section{Results}
The $^{9}$Be($^{28}$Mg,$^{26}$Ne)X inclusive cross section was
determined from the number of detected $^{26}$Ne residues relative to the
number of incoming $^{28}$Mg and target nuclei. The result was $\sigma^\text{inc}= 1.475(18)$~mb, in excellent agreement with the previous measurement
of 1.50(10)~mb \cite{bazin03}. Besides the statistical error, the
quoted uncertainty includes contributions from the acceptance correction,
beam intensity, target thickness, particle identification and detection
efficiency.

\begin{figure}[h]
\centering
\includegraphics[width=\columnwidth]{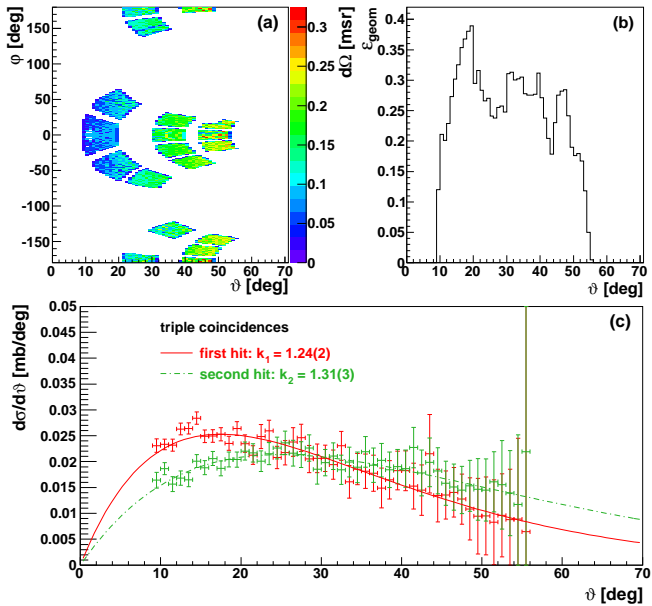}
\caption{(color online) Extrapolation of the measured cross section for the range of polar angles not covered by HiRA. (a) Solid angle covered by the HiRA array in the configuration used in the present study. (b) Geometric efficiency of the HiRA array. (c) measured cross section for triple coincidences as a function of $\vartheta$, where ``first hit'' denotes the detected particle with the lower mass, or the higher energy in case of identical particles (see text).}
\label{fig:extr}
\end{figure}

Triple coincidence events, comprising two charged particles in HiRA and $^{26}$Ne in the spectrograph, were corrected for the geometric efficiency of the HiRA array, i.e. for the azimuthal ($\varphi$) angle coverage, shown in Fig.~\ref{fig:extr} (b). 
The resulting cross section for triple coincidences in the observed polar ($\vartheta$) angular range is $\sigma^\text{tot}_\text{obs}=0.88(2)$~mb. Besides protons also other light charged particles have been detected. 28.4~\%, 51.1~\% and 20.5~\% of the cross section is associated with triple coincidences with two, one and no detected protons in HiRA as shown in Table~\ref{tab:parttypes}. 
\begin{table}[h]
\caption{Triple coincidence cross sections in the polar angular range covered by the HiRA detectors. In addition to proton-proton coincidences (pp), also events where one (px) or both (other) of the detected particles is another light ion (deuteron, triton, $^3$He or $\alpha$ particle) or an unidentified particle, with an energy below the identification threshold, have been observed.}
\begin{ruledtabular}
\begin{tabular}{r|c|c|c}
  & fraction [\%] & $\sigma_\text{obs}$ [mb] & $\sigma_\text{extr}$ [mb] \\
\hline
tot & & 0.88(2) & 1.43(5)\\
\hline
pp  & 28.4 & 0.25(2) & \\ 
\hline
px  & 51.1 & 0.45(4) & \\ 
\hline
other  & 20.5 & 0.18(2) & \\ 

\end{tabular}
\end{ruledtabular}
\label{tab:parttypes}
\end{table}
The observed deuteron, triton, $^3$He or $\alpha$ particles are originating from inelastic (pickup) reactions of the removed protons with the target nucleons. Low energy fragments of the target itself have not been observed.

The triple coincidence cross section ($\sigma^\text{tot}_\text{obs}=0.88(2)$~mb) is then corrected for the range of polar ($\vartheta < 9^\circ$ and $>56^\circ$) angles not covered by HiRA (see Fig.~\ref{fig:extr} (b)).
The angular distribution $d\sigma/d\vartheta$ of light particles for triple coincidences is shown in Fig.~\ref{fig:extr} (c).
Here the first-hit distribution refers to the detected particle in HiRA with the smaller mass (i.e. the proton in case of proton-deuteron-residue coincidences), or, in the case of two identical light particles detected, to the particle with the higher energy. The second hit denotes the particle detected in HiRA with the larger mass, or lower energy.
These angular distributions are then fitted with an exponential function combined with the solid angle factor $2\pi\sin\vartheta$ to extract extrapolation parameters $k_{1,2}$ to correct for the unobserved cross section. Extrapolated cross sections are then obtained by multiplying the cross section in the observed polar angular range with the extrapolation parameters, i.e.
\begin{equation}
\sigma_\text{extr}=k_1 \cdot k_2 \cdot \sigma_\text{obs}.
\end{equation}

The extrapolated total cross section is $\sigma^\text{tot}_\text{extr}=1.43(5)$~mb, in excellent agreement with the inclusive cross section.
This means that for essentially every knockout event one finds two light charged particles in the exit channel, originating
either from the diffraction mechanism or from more complex (inelastic)
reactions of the removed protons with the target nucleons.
Similarly, the total cross section can also be extracted independently from events where only one light particle was detected in coincidence with the $^{26}$Ne residue. Here the number of counts had to be divided by two to correct for the fact that two particles were emitted ($\sigma^\prime_\text{obs}=0.85(2)$~mb). Assuming that there is no angular correlation between the two light particles ($k^\prime_1 = k^\prime_2 = 1.31(2)$), the total cross section extracted using this method amounts to $\sigma^\prime_\text{extr}=1.46(4)$~mb in agreement with the inclusive and triple-coincidence-deduced values.

\begin{figure}[!h]
\centering
\includegraphics[width=\columnwidth]{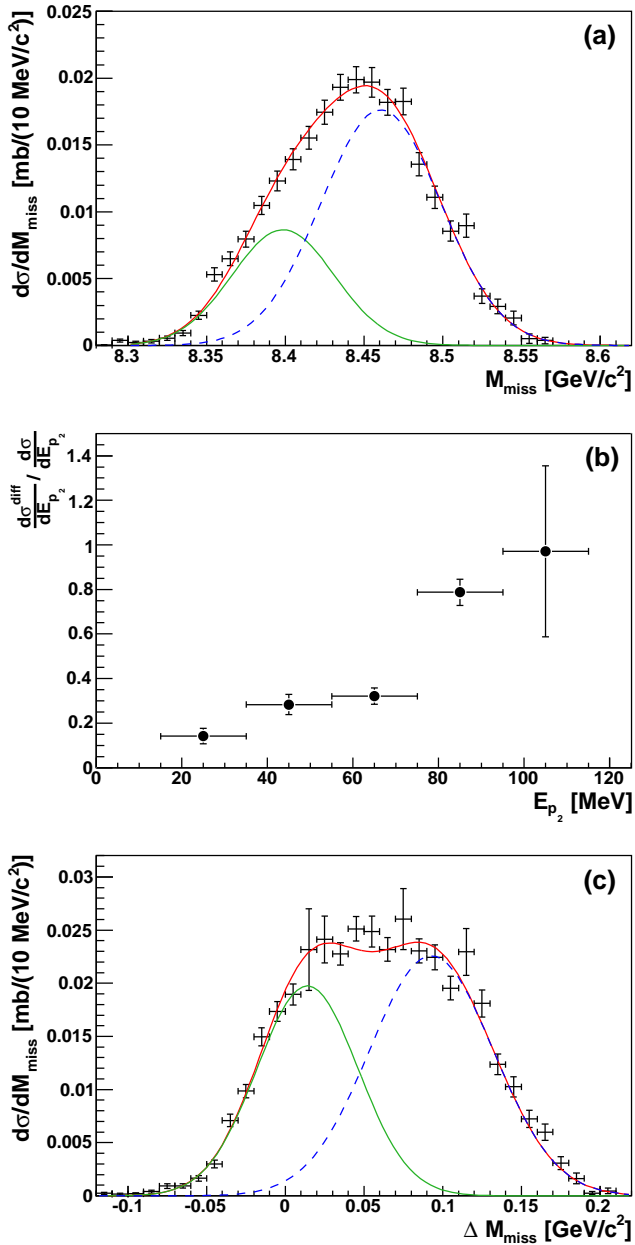}
\caption{(color online) (a) Missing-mass spectrum for events where two
protons were detected. The cross section for observing two protons
in coincidence with the $^{26}$Ne residue amounts to $\sigma^\text{pp}_\text{obs}=0.25(2)$~mb. The spectrum was fitted with two Gaussian peaks. The lower peak, at the target mass, is due to the diffraction mechanism (green, solid line), the larger peak is attributed to events where at least one proton was removed in an inelastic collision with the target. (b) Relative diffraction yield as a function of the smaller of the two proton energies, $E_{\text{p}_2}$. (c) Missing-mass difference spectrum for events where one of the detected particles is a proton, and the other is a light ion, e.g. a deuteron or triton. The individual spectra have been shifted ($\Delta M_\text{miss}=M_\text{miss}-M(^{8}\text{Be}), M(^{7}\text{Be})$ etc. and then added. From the fit the ratio of diffraction-stripping to stripping events was extracted.}
\label{fig:pp_missmass}
\end{figure}

To determine the individual contributions from the three removal
reaction mechanisms, the missing mass of each triple coincidence event was reconstructed
from the observed energies and momenta of the three particles.
Fig.~\ref{fig:pp_missmass} (a) shows the missing-mass spectrum for events
where two protons were detected. 
With this data selection, all three
reaction processes that remove two protons, i.e. stripping,
diffraction-stripping, and diffraction, can contribute.
The spectrum was fitted with two Gaussian functions leaving all six parameters free to vary. The lower peak, at $M_\text{miss}=8.399(3)$~GeV/c$^2$, is at the mass of the target nucleus ($M(^{9}\text{Be})=8.395$~GeV/c$^2$),
and is thus attributed to the diffraction mechanism since no energy is transferred to the target in this elastic two-proton removal process. Its width ($\sigma =32.5(11)$~MeV/c$^2$) is in good agreement
with the expected resolution from the energy spread of the incoming beam ($25$~MeV),
the differential energy loss in the target ($21$~MeV), and the combined resolution
for light particles detected in HiRA ($5$~MeV). The second, broader distribution originates
from the inelastic stripping and diffraction-stripping mechanisms, which
cannot be resolved in this spectrum. The extracted cross section
for diffraction of both protons in the polar angular range covered by HiRA
amounts to $\sigma^\text{diff}_\text{obs}=0.07(2)$~mb.
As is expected for the
diffraction mechanism, its cross section increases for large proton
energies. This is shown in Fig.~\ref{fig:pp_missmass} (b) where the relative
diffraction yield, $(d\sigma^\text{diff}/dE_{\text{p}_2}) / (d\sigma/dE_{\text{p}_2})$, is plotted as a function of the smaller of the two proton
energies $E_{\text{p}_2}$.

The ratio of diffraction-stripping to stripping events was extracted by
fitting similar missing-mass spectra for events where one of the detected particles
is a proton, and the other is a light ion, e.g. a deuteron or triton.
Since the residue nucleus $^{26}$Ne is identified in the S800 spectrograph, the additional neutrons in the light particles can only originate from inelastic pickup reactions on the target.
In a one-proton knockout reaction these pickup reactions lead to a sharp peak in the missing mass at $M_\text{miss}=M(^{8}\text{Be})=7.456$~GeV/c$^2$ in case of residue-deuteron coincidences and $M_\text{miss}=M(^{7}\text{Be})=6.536$~GeV/c$^2$ for events where tritons are detected~\cite{1pknock}. In the two-proton knockout reaction the missing mass is thus expected to be $M_\text{miss}=M(^{7,8}\text{Be})$ for events where the proton was removed in an elastic collision and larger values for $M_\text{miss}$ for reactions where the proton was removed in a stripping reaction. Missing-mass spectra for triple coincidence events with only one identified proton were fitted with two components as shown in Fig.~\ref{fig:pp_missmass} (c) to obtain the ratio of diffraction-stripping to stripping events.
Similar to Fig.~\ref{fig:pp_missmass} (b) the yield of diffraction-stripping
events increases with the proton energy, while the stripping yield $(d\sigma^\text{str}/dE_\text{p}) / (d\sigma/dE_\text{p})$
stays constant as a function of the energy of the detected proton $E_\text{p}$. This feature is
independent of the type of the second detected particle.
The ratio of diffraction-stripping to stripping amounts to
$\sigma_\text{diff-str}/\sigma_\text{str}=0.7(2)$ for
events where one of the two particles is a proton.
This ratio is then used to extract the diffraction-stripping and stripping cross section from the events where both detected light particles were protons, $\sigma^\text{pp}_\text{diff-str+str}=0.17(2)$~mb, the broad component in Fig.~\ref{fig:pp_missmass} (a), assuming that
the removal processes for the two protons are independent.
Triple coincidence events, where none of the detected light particles is a
proton ($\sigma^\text{other}_\text{obs} = 0.18(2)$~mb), are assumed to arise
from events where both protons were stripped. 

The cross sections for the three
mechanisms are summarized and compared to the theoretical predictions
\cite{tostevin06} (recalculated for the present incident beam energy) in
Table~\ref{tab:cs}.
\begin{table}[h]
\caption{Cross sections of the three mechanisms contributing to the
two-proton removal reaction and comparison with theory. For the
comparison of the cross sections the theoretical values have been
multiplied with the factor $R_\text{S}(2\text{N})=0.488(6)$. The 
relative contribution of each mechanism is consistent with theory.}
\begin{ruledtabular}
\begin{tabular}{r|c|c|c|c}
  & diff & diff-str & str & tot.\\
\hline
$\sigma_\text{obs}$~[mb]                   & 0.07(2) & 0.27(14) & 0.54(14) & 0.88(2)  \\
$\sigma_\text{extr}$~[mb]                  & 0.11(3) & 0.44(23) & 0.87(23) & 1.43(5)  \\
\hline
fraction [\%]                              & 8(2)    & 31(16)   & 61(16)   &          \\
\hline
$\sigma^\text{inc}$ [mb]                   &         &          &          & 1.475(18)\\
\hline
$\sigma_\text{theo}$ incl. [mb]            & 0.19    & 1.13     & 1.70     & 3.02     \\
$\sigma_\text{theo}\cdot R_\text{S}(2\text{N})$ [mb] & 0.09     & 0.55     & 0.83     & 1.475   \\
\hline
fraction$_\text{theo}$ [\%]                & 6.3     & 37.4     & 56.3     &          \\
\end{tabular}
\end{ruledtabular}
\label{tab:cs}
\end{table}
The agreement with the theoretical expectations for the relative importance
of each contribution is very good. 
Comparison of the absolute cross section values confirms
the need for a reduction factor $R_\text{S}(2\text{N})=\sigma_\text{exp}/
\sigma_\text{theo} = 0.488(6)$, in agreement with the value $R_\text{S}(2\text{N})=0.50(3)$~\cite{tostevin06} deduced from
the partial and inclusive cross sections reported in the previous
$^{26}$Ne--$\gamma$ coincidence experiment\cite{bazin03}.

\section{Summary}
In summary, first coincidence measurements of fast, light charged 
particles and the heavy projectile-like reaction residues following 
the removal of two well-bound protons from $^{28}$Mg, have allowed 
the relative importance of the stripping, diffraction-stripping and 
diffraction two-proton removal mechanisms to be determined. The 
experimental results are consistent with the expectations for the 
relative importance of each mechanism calculated using the eikonal 
reaction framework~\cite{tostevin06} that assumes a sudden, one-step 
removal of the two protons from the fast projectile. The measured 
cross sections thus confront these specific reaction-model predictions
\cite{tostevin06,simpson09} at a more detailed level. The measurements 
are also consistent with the previously-determined reduction factor 
$R_\text{S}(2\text{N})$ for two-proton knockout from $^{28}$Mg and 
for two-nucleon removals from other $sd$-shell systems. These results 
further support the applicability and effectiveness of this reaction 
channel, that populates residues even more exotic than the projectile, 
for the production of and extraction of spectroscopic information 
in some of the most exotic nuclei currently accessible.

This work was supported by the National Science
Foundation under Grant No. PHY-0606007 and PHY-0757678, the DOE/NNSA
(National Nuclear Security Administration) Grant No. DE-FG55-08NA28552,
and the United Kingdom Science and Technology Facilities
Council (STFC) under Grant No. ST/012012/1 and ST/J000051/1.

\bibliography{28mg_2p}

\end{document}